\documentclass{article}
\usepackage[T1]{fontenc} % add special characters (e.g., umlaute)
\usepackage[utf8]{inputenc} % set utf-8 as default input encoding
\usepackage{ismir,amsmath,cite,url}

\usepackage{graphicx}
\usepackage{color}
\usepackage{booktabs} 
\title{The Bach Doodle: Approachable music composition with machine learning at scale}

\authorblurb {
\begin{tabular}{@{}c@{}}
\parbox{\dimexpr\linewidth-2\fboxsep-2\fboxrule}{
\centering
{\bf
Cheng-Zhi Anna Huang$^\sharp$\quad
Curtis Hawthorne$^\sharp$\quad
Adam Roberts$^\sharp$\quad
Monica Dinculescu$^\sharp$\quad
James Wexler$^\dagger$\quad
Leon Hong$^\star$\quad
Jacob Howcroft$^\star$
} \\
{\tt \{annahuang,fjord,adarob,noms,jwexler,leonhong,jhowcroft\}@google.com} \\
{\small $\sharp$ Google Brain Magenta \quad $\dagger$ Google Brain Pair \quad $\star$ Google Doodle}
}
\end{tabular}
}

\def\authorname{
    {Cheng-Zhi Anna Huang},
    {Curtis Hawthorne},
    {Adam Roberts},
    {Monica Dinculescu},
    {James Wexler},
    {Leon Hong},
    {Jacob Howcroft}}

\newcommand{\vect}[1]{\mathbf{#1}}
\newcommand{\given}{~|~}

\newcommand{\ctx}{C}

\begin{document}

\maketitle
\section{Abstract}

%To celebrate Johann Sebastian Bach's 334th birthday, we wanted everybody to experience the joy of writing music.

%Composing traditionally has a high-barrier to entry, requires specialized skills, even for musicians it might feel hard to come up with good ideas.

To make music composition more approachable, we designed the first AI-powered Google Doodle, the Bach Doodle~\cite{doodle}, where users can create their own melody and have it harmonized by a machine learning model (Coconet~\cite{huang2017counterpoint}) in the style of Bach.  
For users to input melodies, we designed a simplified sheet-music based interface.
To support an interactive experience at scale, we re-implemented Coconet in TensorFlow.js~\cite{smilkov2019tensorflow} to run in the browser and reduced its runtime from $40$s to $2$s by adopting dilated depth-wise separable convolutions and fusing operations.  
We also reduced the model download size to approximately $400$KB through post-training weight quantization. 
We calibrated a speed test based on partial model evaluation time to determine if the harmonization request should be performed locally or sent to remote TPU servers.
In three days, people spent 350 years worth of time playing with the Bach Doodle, and Coconet received more than 55 million queries.
Users could choose to rate their compositions and contribute them to a public dataset, which we are releasing with this paper.  We hope that the community finds this dataset useful for applications ranging from ethnomusicological studies, to music education, to improving machine learning models.  

\section{Introduction}
%Many of us enjoy music, but composing can feel intimidating.  Even when we have a melody, without sufficient skills in harmony we are deterred from developing it into a composition.  
%Even with the technical skills, it is often time consuming to work out the details of a harmonization and it slows us down from iterating.

Machine learning can extend our creative abilities by offering generative models that can rapidly fill in missing parts of our composition, allowing us to see a prototype of how a piece could sound.  
To celebrate J.S. Bach's 334th birthday, we designed the Bach Doodle to create an interactive experience where users can rapidly explore different possibilities in harmonization by tweaking their melody and requesting new harmonizations.  The harmonizations are powered by Coconet~\cite{huang2017counterpoint}, a versatile generative model of counterpoint that can fill in arbitrarily incomplete scores.

Creating this first AI-powered doodle involved overcoming challenges in user interaction and interface design, and also technical challenges in both machine learning and in infrastructure for serving the models at scale.
For inputting melodies, we designed a simplified sheet music interface that facilitates easy trial and error and found that users adapted to it quickly even when they were not familiar with western music notation.  
%Several technical challenges had to be overcome to support this interactive experience at scale.  

As most users do not have dedicated hardware to run machine learning models, we re-implemented Coconet in TensorFlow.js~\cite{smilkov2019tensorflow} so that it could run in the browser.  
We reduced the model run-time from 40s to 2s by adopting dilated depth-wise separable convolutions and fusing operations, and we reduced the model download size to ${\sim}400$KB through post-training weight quantization. 
%our initial re-implementation took more than 40 seconds to generate two measures of music.  By adopting dilated depth-wise separable convolutions and model quantization, we reduced it down to 2 seconds.  
To prepare for large-scale deployment, we calibrated a speed test to determine if a user’s device is fast enough for running the model in the browser. If not, the harmonization requests were sent to remote TPU servers.

%To support an interactive experience at scale, we re-implemented Coconet in TensorFlow.js~\cite{smilkov2019tensorflow} to run in the browser and reduced its runtime from 40s to 2s by adopting dilated depth-wise separable convolutions and model quantization.  We calibrated a speed test based on partial model evaluation time to determine if the harmonization request is performed locally or sent to remote TPU servers.

%Even though music notation can be intimidating for novices, we show that through an intuitive and simplified sheet music interface

%facilitates easy trial and error, and found that users adapted to it quickly even when they were not familiar with western music notation.

%assisted by machine learning, composing Bach-chorale style counterpoint can become more approachable. 

Users in 80\% of sessions explored multiple harmonizations, and 53.8\% of the harmonizations were rated as positive.  One complaint from advance users was the presence of parallel fifths (P5s) and octaves (P8s).  We analyzed 21.8 million harmonizations and found that P5s and P8s occur on average 0.365/measure and 0.391/measure respectively.  Furthermore, P5s and P8s were more common when user input was out of distribution, and fewer P5s and P8s were correlated with positive user feedback.

\section{Related work}\label{sec:introduction}

Machine learning has been used in algorithmic music composition to support a wide range of musical tasks~\cite{papadopoulos1999ai,fernandez2013ai,pasquier2016introduction,briot2017deep,herremans2017functional}.  
%and its relevance in musical practice discussed~\cite{sturm2019machine}.  
%
Melody harmonization is one of the canonical tasks~\cite{farbood2001analysis, pachet2001musical, chuan2007hybrid, herremans2013composing}, encourages human-computer interaction~\cite{pachet2003continuator,assayag2006omax,fiebrink2011real,huang2018mixed,sturm2019machine}, and is particularly approachable for novices.
Different interfaces and tools have been developed to make the interaction experience more accessible.  For example, in MySong~\cite{simon2008mysong}, users can sing a melody and have the system harmonize it.  In Hyperscore~\cite{farbood2004hyperscore}, users can draw multiple levels of ``motifs'' on a graphical sketchpad and have them harmonized according to the tension curve they specified. 
Startups such as JukeDeck and Amper Music offer APIs that allow users to describe a piece through timing and mood tags. %~\cite{dredge2019music}.  
Opensource libraries such as Magenta.js~\cite{roberts2018magenta} allow machine learning models to be used in digital audio work stations. 
For score-based interaction, FlowComposer~\cite{papadopoulos2016assisted} offers an augmented lead-sheet based interface, while DeepBach~\cite{hadjeres2016deepbach} demonstrates interactive chorale composition in MuseScore which uses standard music notation.

\section{The Bach Doodle}

\subsection{A walk through of the user experience}
The Bach Doodle user experience begins by demonstrating 4-part harmony using two measures of a Bach chorale, \textit{Ach wie flüchtig, ach wie nichtig, BWV 26}. By playing the soprano line alone, followed by soprano and alto, and then all four voices, users are shown how the harmony enhances the melody.
Users are then presented with two measures of blank sheet music, in the treble clef, with a key signature of C major, in standard time. There are four vertical lines in each measure to indicate the beats which give the user visual cues on where to put notes.

\begin{figure}[h]
\centering
\includegraphics[width=\columnwidth]{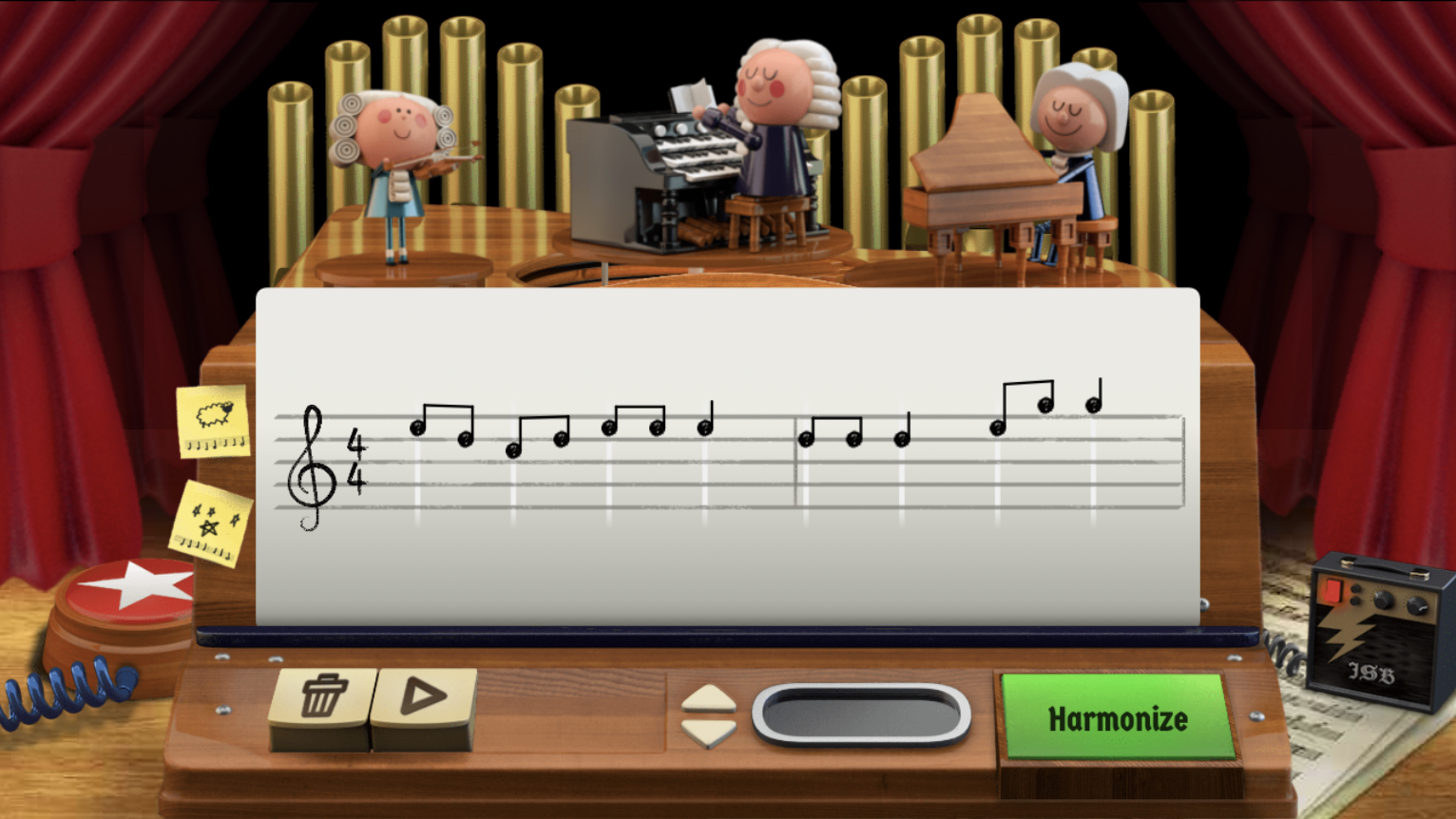}
\caption{The user interface of the Bach Doodle, where users can input a melody and then click on the green ``Harmonize'' button on the bottom right to request a harmonization.}
\label{fig:bach_doodle_melody}
\end{figure}

The user enters a monophonic melody using quarter and eighth notes. The note duration is automatic. If a note is entered on a beat, it is a quarter note by default. However, if a note is added after the beat, the existing quarter note becomes an eighth note. This simple interface makes it easy for users with no musical knowledge to input a composition, and can be seen in Figure~\ref{fig:bach_doodle_melody}.
If the user clicks on the ``star'' button on the left-hand side of the sheet music, they can enter ``advanced mode''. This allows the user to input sixteenth notes anywhere on the staff. It also enables a control where the key can be changed to any of the 12 key signatures. This mode is hidden because it can be overwhelming for new users. It is also easier to make less enjoyable music this way, for example by going off key, or making the music overly complex.

\begin{figure}[h]
%\begin{wrapfigure}{l}{0.60\textwidth}
\centering
\includegraphics[width=\columnwidth]{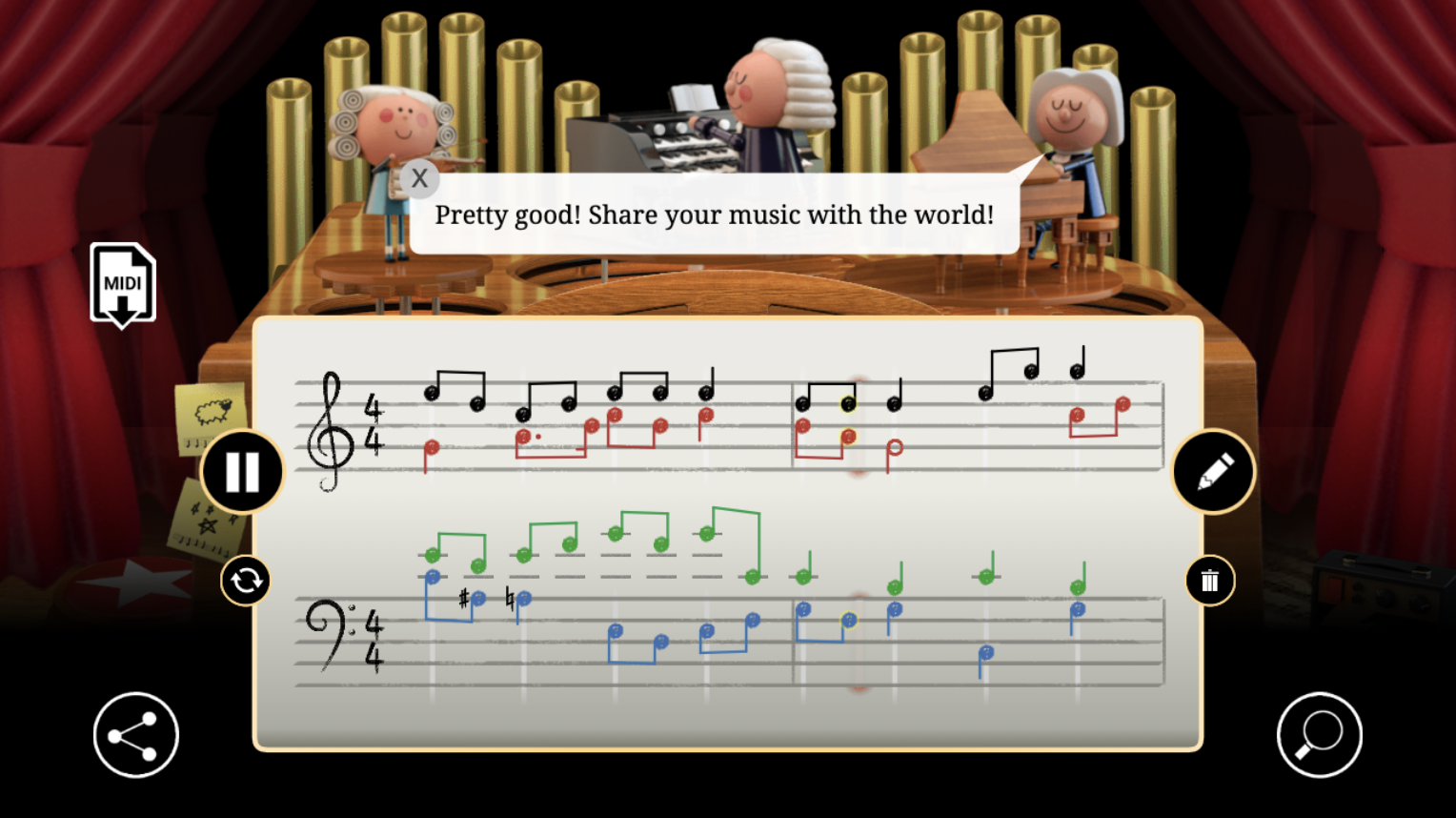}
\caption{The harmonization returned by Coconet is notated in color, carrying the alto, tenor, and bass voices.}
\label{fig:bach_doodle_harm}
\end{figure}

Clicking the ``Harmonize'' button sends the generation request to either TensorFlow.js or the TPU server. When the response is ready, it is presented to the user one voice at a time, listing them out: ``Soprano'', ``Alto'', ``Tenor'', ``Bass''. The voices are color-coded to illustrate the harmonization in relation to the soprano input notes (Figure~\ref{fig:bach_doodle_harm}).

%\begin{wrapfigure}{l}{0.60\textwidth}
%\end{wrapfigure}

%\end{wrapfigure}

\subsection{Design challenges}
Celebrating J.S. Bach's birthday using machine learning presented many unique design opportunities as well as some user experience challenges. One of the main goals was to empower people with the feeling that they could augment their own creativity in ways not previously thought possible, by allowing them to directly collaborate with a machine learning model. Another important goal was to convey the message that machine learning is not “magical” or incomprehensible, but rather a science that can be understood. Finally, a notable challenge was to ensure that these aims would be met for a large diversity of individuals, from children who have not yet learned to read to experts of music and technology.

In order to acquire early feedback on the design, user tests were employed. Over the course of the project, dozens of people were asked to play the demos and comment on their experience through both pre-defined questions and open comments. The first user test in the development process revealed that many people do not fully understand the concept of harmony, but fortunately, further testing showed that short animated musical examples were enough for people to comprehend these concepts quickly. Also, user tests indicated that only a small subset of people could read standard music notation. Our intuition was that using standard notation, rather than a grid based interface would be intuitive and frictionless to anybody only if the user interface (UI) was responsive with animations and sound and also if the note input interface was kept simple. Further user testing of the standard notation input UI proved this to be correct.

In order to accommodate people of all ages and experiences, a common technique employed is to remove any advanced feature or unexpected delight from the core experience and instead integrate them as ``easter eggs''. This allows people of all skill levels to experience the full core experience without feeling frustrated, while also giving the rarer advanced user more features. While the core experience primarily allows eighth notes and tempo changes, clicking on a special button in the background additionally allowed the user to add sixteenth notes and change the key -- two features that are very confusing to those without musical backgrounds.

\subsection{Reusable insights}
For future projects, we have shown that if the technology being used is unfamiliar or perceived as “scary” to those who know little about it, tethering the experience to a familiar story and visuals can be a successful strategy. Most people have a limited understanding of musical concepts such as harmony and standard notation, but it is possible that people of all ages can quickly acquire an intuitive understanding of musical concepts through carefully designed animated audiovisuals and a simple and responsive UI. Additionally, injecting content into loading screens could not only make loading times feel shorter but also be an excellent space for educational content. Finally, user testing is crucial when trying to create an experience using new technology that encompasses a large and diverse audience -- it can reveal serious issues and shortcomings that are not obvious due to the team’s own background and domain knowledge.

% \section{The launch}
% First launch is 3/20 at 8am Pacific. Last un-launch is 3/22 at 11pm Pacific
% in how many countries?

\section{Technical challenges}

In order for users to interact with Coconet via a web interface, we needed to either port it to run client-side on the user's device or host the model on a server with sufficient speed and capacity to support the number of requests we were expecting. In fact, we did both: we ported the model to TensorFlow.js (TF.js) so that it could run on user devices and added support for Tensor Processing Units (TPU) so that it could be served on Google Cloud. By running a simple test on users' devices to determine the speed of core tensor operations, we were able to determine whether to use the TF.js implementation locally, or fall back to a TPU server to handle the computation. In the end, 47.4\% of all harmonizations were done locally, in TF.js.

\subsection{Background: Coconet}
\begin{figure}[h]
  \includegraphics[width=\linewidth,trim = 0.0cm 1.5cm 0.0cm 0.5cm]{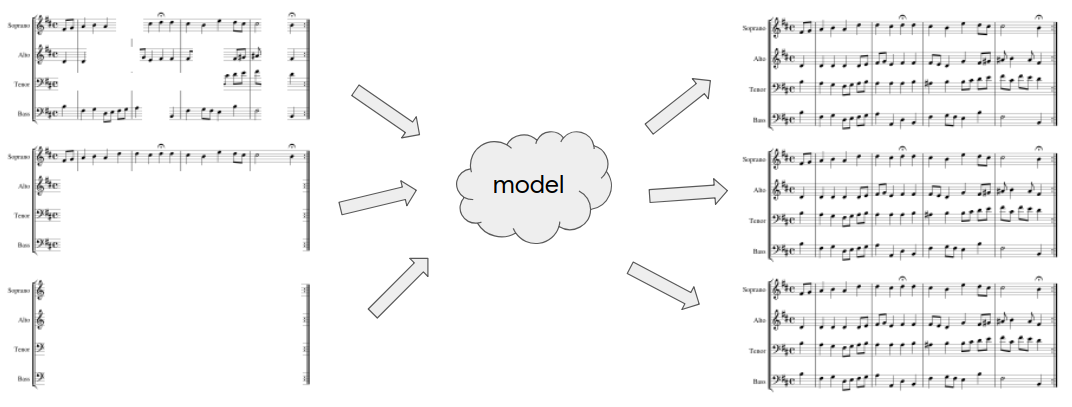}
  \caption{Coconet can be used in a wide range of musical tasks, such as completing partial scores, harmonizing melodies and generating from scratch.}
  \label{fig:coconet-task}
\end{figure}

Coconet~\cite{huang2017counterpoint}~\footnote{Blog post: \url{https://magenta.tensorflow.org/coconet} \newline Code: \url{https://github.com/tensorflow/magenta/tree/master/magenta/models/coconet}} is a versatile generative model of musical counterpoint that can fill in arbitrarily incomplete scores, as illustrated in Figure~\ref{fig:coconet-task}. 

Coconet represents counterpoint as a stack of piano rolls encoded in a binary three-dimensional tensor $\vect{x} \in \{0, 1\}^{I \times T \times P}$, where $I$, $T$, and $P$ denotes the number of instruments, time steps, and pitches respectively.  
$\vect{x}_{i,t,p} = 1$ if the $i$th instrument plays pitch $p$ at time $t$.
Each instrument is assumed to play exactly one pitch at a time, therefore $\sum_p \vect{x}_{i,t,p} = 1$ for all $(i, t)$ positions.
We also focus on four-part Bach chorales as used in prior work~\cite{cope1991computers,allan2005harmonising,boulanger2012modeling,goel2014polyphonic,liang2016bachbot,hadjeres2016style}, and assume $I = 4$ throughout.
%We constrain ourselves to only the range that appears in our training data (MIDI pitches 36 through 88).

\begin{figure}[h]
  \includegraphics[width=\linewidth, trim = 0.0cm 2.0cm 0.0cm 2.5cm]{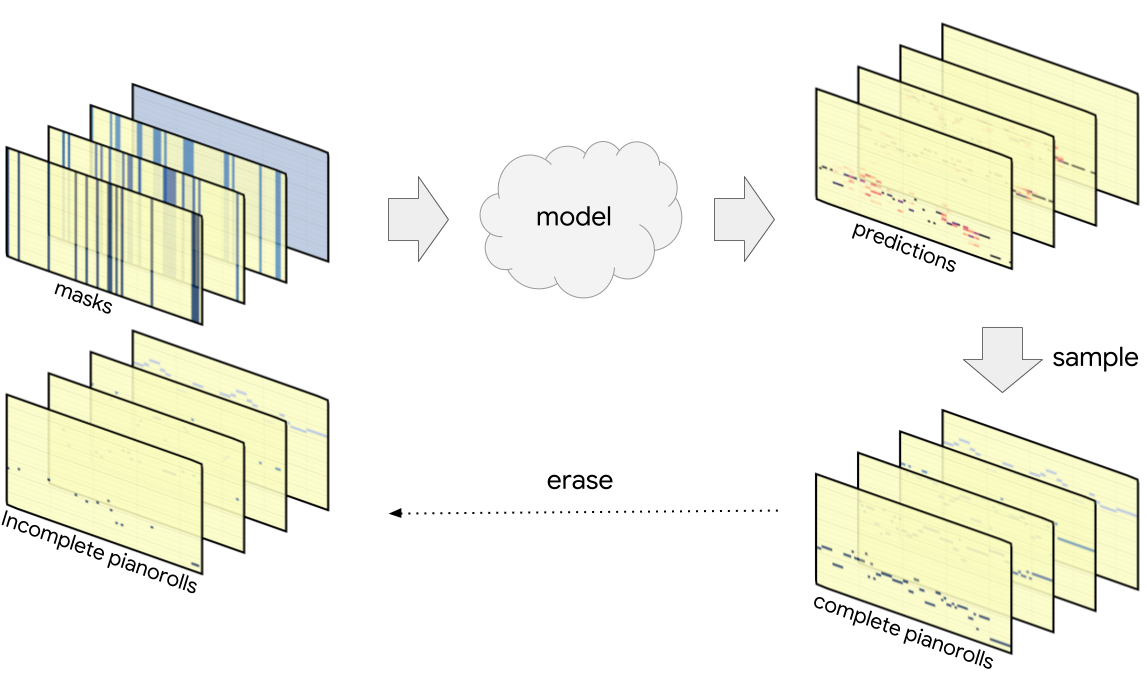}
  \caption{Coconet's generation loop using Gibbs sampling, alternating between (top) filling in the missing parts and (bottom) erasing random parts to improve the score through rewriting.}
  \label{fig:coconet}
\end{figure}

Conventional approaches often factorize the joint distribution $p(\vect{x})$ into conditional distributions of the form $p(\vect{x}_k \given \vect{x}_{<k})$, where $k$ indexes a sequence in some predetermined ordering such as chronological. In contrast, Coconet is an instance of orderless NADE~\cite{uria2014deep, uria2016neural} and offers direct access to all conditionals of the form
$p(\vect{x}_{i,t} \given \vect{x}_\ctx)$, where
$\ctx$ selects a fragment of a musical score $\vect{x}$ and $(i, t) \in \neg\ctx$ is in its complement (i.e. the missing parts). 
To train Coconet, we sample a training example $\vect{x}$, 
choose uniformly how many variables to erase, i.e. $|\neg\ctx| \sim U(1, D)$, and then choose uniformly the particular subset of variables $\neg\ctx$ to erase. 
The input $\mathbf{X} \in \{0,1\}^{2I \times T \times P}$ is obtained by erasing the piano rolls $\vect{x}$
to obtain incomplete piano rolls $\vect{x}_\ctx$ 
and concatenating this with the corresponding masks, as shown in Figure~\ref{fig:coconet} (top left) where the yellow gaps indicate erased positions with all pitches set to zero.

% The Gibbs sampling loop which Coconet completes partial scores using Gibbs sampling, by alternating between filling in incomplete scores and randomly erasing parts of the score so that it can be better filled in given the current context.   to iterate between  sampling to fill in missing parts.  (Left) The input is concatenation of incomplete pianorolls and a set of masks that indicate where notes are missing.  The model produces predictions, which are softmax distributions over pitch for each instrument / timestep pair.  These distributions are sampled independently to produce

% The input $\vect{h}^0 \in \{0,1\}^{2I \times T \times P}$ is obtained by masking the piano rolls $\vect{x}$
% to obtain the context $\vect{x}_\ctx$
% and concatenating this with the corresponding mask with $0$ indicating erased and $1$ indicating kept:
% \begin{align}
% \vect{h}^{0}_{i,t,p} &= \mathbbm{1}_{(i, t) \in \ctx} \vect{x}_{i,t,p} \\
% \vect{h}^{0}_{I + i,t,p} &= \mathbbm{1}_{(i, t) \in \ctx}
% \end{align}
%
% The final activations $\vect{h}^L \in \mathbb{R}^{I \times T \times P}$ are passed through the softmax function to obtain predictions for the pitch at each instrument/time pair:

% \begin{align}
% \label{eqn:partial_loss}
% p_\theta(\vect{x}_{i,t,p} \given \vect{x}_\ctx, \ctx) =
% \frac{\exp(h^L_{i,t,p})}{\sum_p \exp(h^L_{i,t,p})}
% \end{align}

The output predictions for each $(i, t)$ position is a softmax over the set of pitches $P$ (top right of Figure~\ref{fig:coconet}).  The negative loglikelihood loss is given below,
%in Equation~\ref{eqn:partial_loss}, 
which involves reweighing by the number of variables erased to ensure that all conditionals are trained equally.
%\label{eqn:partial_loss}
\begin{align}
\mathcal{L}(\vect{x}; \ctx)
%   &= - \frac{1}{|\neg\ctx|} \sum_{(i, t) \notin \ctx}
%    \log p_\theta(\vect{x}_{i,t} \given \vect{x}_\ctx, \ctx)
   &= - \frac{1}{|\neg\ctx|} \sum_{(i,t) \in \neg\ctx} \sum_p \vect{x}_{i,t,p}
       \log p(\vect{x}_{i,t,p} \given \vect{x}_\ctx, \ctx) \notag
\end{align}

In contrast to generating from left to right in one pass, Coconet uses Gibbs sampling to improve sample quality through rewriting (see~\cite{huang2017counterpoint} for convergence analysis).  Figure~\ref{fig:coconet} shows how the procedure iterates between filling in missing parts and then erasing other parts so that they could be rewritten given the updated context.

\subsection{Improving and speeding up Coconet}
The original Coconet uses dense convolutions, where filter weights and their mixing weights $W$ are fully connected (Equation~\ref{eqn:dense_conv}). This makes it unable to fully leverage GPU parallelization in TF.js (see Section~\ref{sec:performance_challenges}).   

Let $s, q$ index the pitch and time dimension in filters, and $i, j$ the input and output channels.  In dense convolutions, each output position $(p, t, j)$ indexed by pitch, time and output channel is a sum over the resultant input channels and also over the positions in each filter (given by the $\text{neighbourhood}$ function). For a 3-by-3 filter this summing is over the 9 positions.

In contrast, depthwise separable convolutions~\cite{chollet2017xception}, shown in Equation~\ref{eqn:dsep_conv}, factorizes the dense tensor $W$ into a depthwise tensor $V$ and a pointwise $U$. As a result, the multiplications between $V$ and $X$ can be parallelized over the input channels $i$ in the inner sum of Equation~\ref{eqn:dsep_conv_2}.
%of size 3 by 3. 
%
\begin{align}
Y^{\text{dense}}_{p, t, j} &= \sum_i\sum_{s, q \in \text{neighborhood(p, t)}} W_{s, q, i, j} X_{s, q, i} \label{eqn:dense_conv}\\
Y^{\text{dsep}}_{p, t, j} &= \sum_i
    \sum_{s, q \in \text{neighborhood(p,t)}} U_{i, j}V_{s, q, i} X_{s, q, i} \label{eqn:dsep_conv} \\
                         &= \sum_i U_{i, j}
    \sum_{s, q \in \text{neighborhood(p,t)}} V_{s, q, i} X_{s, q, i} \label{eqn:dsep_conv_2}
\end{align}
To further speed up Coconet, we adopt dilated convolutions to grow the receptive field exponentially to reduce the number of layers needed.  As in \cite{oord2016wavenet}, where in each block the dilation factors double in each layer for both the pitch and time dimension and then the block repeats. 

The original Coconet was trained on eight measures (T=128).  However, the Bach Doodle is designed for two measures (T=32), so we retrained the model with the original architecture and saw that the loss increased from 0.57 to 0.62 (show in Table~\ref{table:likelihood}), possibly because there is less context.  Switching from dense to depthwise separable convolutions reduced the loss, requiring more filters but fewer layers.  Since Tensorflow.js allows for parallelization across filters, this still resulted in much faster generation (see Section~\ref{sec:performance_challenges}).  Dilated convolutions reduced both the number of layers and number of filters and also reducing the loss. The particular scheme we used is 7 blocks of dilation rates $(1, 2, 4, 8, 16, 16)$ for the pitch dimension and $(1, 2, 4, 8, 16, 32)$ for the time dimension.

\begin{table}[h]
  \caption{Comparing frame-wise negative loglikelihood (NLL) on the 16th-note resolution as in~\cite{huang2017counterpoint} and the generation time (in seconds) when model was ported to Tensorflow.js (see Section~\ref{sec:performance_challenges} for details). The bottom three rows are all trained on two-measure (T=32) random crops.}
  \label{table:likelihood}
  \centering
  \begin{tabular}{lrr}
    \toprule
    Convolution type & NLL & run time\\
    \midrule
    Dense (T=128), 64L, 128f & 0.57\\  
    \midrule
    Dense (T=32), 64L, 128f & 0.62 & > 40s \\
    % train-test-combined.sepconv-numlayersfilters.l48-f256-dm1-rcl0 0.576711473528
    % https://colab.corp.google.com/drive/1Dr_KOwEhMhBipZ-F_BSKnNLR0fBWUNqd?authuser=1#scrollTo=QT6NrVfmTipY
    
    % https://colab.corp.google.com/drive/1Dr_KOwEhMhBipZ-F_BSKnNLR0fBWUNqd?authuser=1
    Depthwise separable, 48L, 192f &  0.59 & 7s\\
    % maxdilated-sepconv.dblocks7-f128-rldlfalse
    % https://colab.corp.google.com/drive/1uTQUXjlhMAM33bJmLexTJPv8AUz6unzj?authuser=1#scrollTo=QT6NrVfmTipY
    % For now, the default network is still the non-dilated winner (48 layers 196 filters).
% The non-dilated models still use 64 steps, but the dilated ones use 96 steps, as suggested during the meeting. Even with the additional steps, the dilated networks are ~25% faster in tfjs than the 48/196 model, which is awesome.
    Dilated, 45L (7 blocks), 128f & 0.58 & $\sim$4s\\ 
    \bottomrule
  \end{tabular}
\end{table}

\subsection{Porting Coconet to the Browser}

JavaScript is the standard language for browser-based computation, but native JavaScript is too inefficient to handle the the amount of computation required by Coconet in a reasonable time for the interaction we desired. TF.js is a javascript library for GPU-accelerated machine learning. It makes use of WebGL~\footnote{\url{https://developer.mozilla.org/en-US/docs/Web/API/WebGL_API}} to leverage the parallel processing power of GPUs to speed up machine learning operations, supporting the development and training of models, as well as deployment of trained models on web browsers. By enabling users to run trained models directly in their web browsers, it alleviates the need for remote servers to run those models. This can enable faster, more interactive experiences between a user and a machine learning system.

While some models can easily be ported to TF.js using a conversion script, Coconet's Python TensorFlow implementation used some ops that did not yet exist in TF.js (e.g., \texttt{cumsum}), and we also needed the flexibility to optimize the performance of the model for our use case. We therefore manually re-implemented Coconet in TF.js ourselves and have made the code opensource\footnote{\url{https://tensorflow.github.io/magenta-js/music/classes/_coconet_model_.coconet.html}}. We also contributed missing ops to TF.js with WebGL fragment shader code for GPU acceleration.

\subsubsection{UI Challenges}

TF.js makes use of the async/await pattern for access to outputs of models and individual TensorFlow operations. During inference, users receive a callback for when GPU operations have completed and the result is ready to be consumed. In this way, there is no blocking of the UI while waiting for model results. In practice, with large models like Coconet (which includes many repeated sampling steps of a deep network), it is still important to cede control back to the UI explicitly during the course of the model operations, which can be done with the \texttt{tf.nextFrame()} operation. Our op-by-op code port of the network allowed us to add these occasional UI breaks, which avoided a poor user experience where the page would freeze for multiple seconds during model prediction. 

\subsubsection{Performance Challenges}
\label{sec:performance_challenges}
The initial port of Coconet to TF.js took over 40 seconds to do one harmonization. For a satisfying user experience, we needed to lessen this latency to below 5 seconds. 
While TF.js is able to take advantage of GPU acceleration, WebGL does not directly support the types of tensor operations used in deep learning. Instead, these operations must be implemented as shader programs, which were originally intended to compute the color for pixels during graphics rendering. This mismatch leads to inefficiencies that sometimes vary by operation. It turned out that the (unavoidably) inefficient shader implementation of convolutional layers were the main culprit. By switching to depthwise-separable convolutions, however, we were able to avoid many of these performance issues, reducing generation time to 7 seconds. 
%
%The negative log-likelihood scores for depthwise-separable convolutional networks were slightly higher than regular convolutions (0.61 vs 0.57 in one test) but still acceptable, and the samples sounded as pleasing as the original models to our ears.
%
%We further reduced the inference latency simply by scaling down the model. We experimented with various reductions in model depth (number of convolutional layers) and width (convolutional filter size), as latency scaled linearly with the number of convolutions, and selected the smallest model that we felt produced compelling harmonizations.

As we run Coconet through 64 Gibbs sampling steps, any improvement to operations that are used in this loop could lead to a significant saving.  We wrote a custom operation using WebGL shaders to fuse together the operations used in our initial TF.js implementation of the annealing schedule. This schedule by \cite{yao2014equivalence} is a sequence of simple element-by-element operations that is run on every sampling step during harmonization. Because of the simplicity of the operations (a scalar subtraction, multiplication, division, and a max operation), we were able to easily fuse them into a single operation that avoided the overhead of executing multiple shader programs on the GPU, speeding up inference by about 5\%.
The combined savings of adding depthwise-separable convolutions, shrinking the model by using dilated convolutions, and using the fused schedule operation resulted in a reduction of the model latency from 40s to 2s.

\subsubsection{Download Size}

Due to the number of users we intended to reach as well as the variety of locations, devices, and bandwidth limits they would have, we needed to ensure the download size of the model weights was as small possible. To achieve this goal, we implemented support for post-training weight quantization and contributed it to TF.js. This quantization compresses each \texttt{float32} weight tensor by mapping the full range of its dimensions down to 256 uniformly-spaced buckets in order to represent them as \texttt{int8} values, which are then stored along with \texttt{float32} min and scale values used to recover the range. During model initialization, the weights are converted back to \texttt{float32} tensors using linear interpolation. By using these quantization, we were able to reduce the size of the downloaded weights by approximately 4, resulting in a payload of $\sim400$KB without any noticable sacrifice in quality.

\subsection{Balancing Load Between Tensorflow.js and TPU}
We ideally wanted to run the harmonization model completely on end-user devices using TF.js to avoid the need for serving infrastructure, which adds cost, effort, and additional points of failure. But the speed of harmonization differs by user device, with older and lower-end devices taking longer to run the TF.js model code. For devices where harmonization take more than a few seconds, the harmonization is instead done by the cloud-served model.
The first step in checking if a device can run harmonization locally is to check if WebGL is supported on the device, since that is required for using GPU-acceleration through TF.js. If WebGL is supported then we perform a speed test on the model, running a sample melody through its first four layers. If the latency of this model inference is below a set threshold, then the TF.js version is used. As there is overhead on the first inference of a model in TF.js, due to initial loading of the model weight textures onto the GPU, we actually run the speed test twice and use the second measurement to make the decision.

\section{Dataset Release and Analysis}
\subsection{Data structure}
Every user who interacted with the Bach Doodle had the opportunity to add their composition to a dataset. We make this entire dataset available at {\url{https://g.co/magenta/bach-doodle-dataset}} under a Creative Commons license.
Of more than 55 million requests, the user contributed dataset contains over 21.6 million miniature compositions.  
%78.8 years were spent composing.  
The compositions are split across 8.5 million sessions.  Each session represents an anonymous user's interaction with the Bach Doodle over a single pageload and may contain multiple data points. Each data point consists of the user’s input melody, the 4-voice harmonization returned by Coconet, as well as other metadata: the country of origin, the user’s rating, the composition’s key signature, how long it took to compose the melody, and the number of times the composition was listened to. 

\subsection{Analysis}
We present some preliminary analysis of the dataset to shed some light on how users interacted with the doodle.  Out of the 21.6 million sequences in the dataset, about 14 million (or 65.7\%) are unique pitch sequences, that are not repeated anywhere in the dataset (without considering timing information). Overall, the median amount of time spent composing a sequence was 25.5 seconds, and sequences were listened to for a median of 3 loops, with a total of 78.2 million loops listened across the entire dataset. 

\begin{figure}[h]
\centering
  \includegraphics[width=0.65\linewidth, trim = 0.0cm 0.7cm 0.0cm 0.0cm]{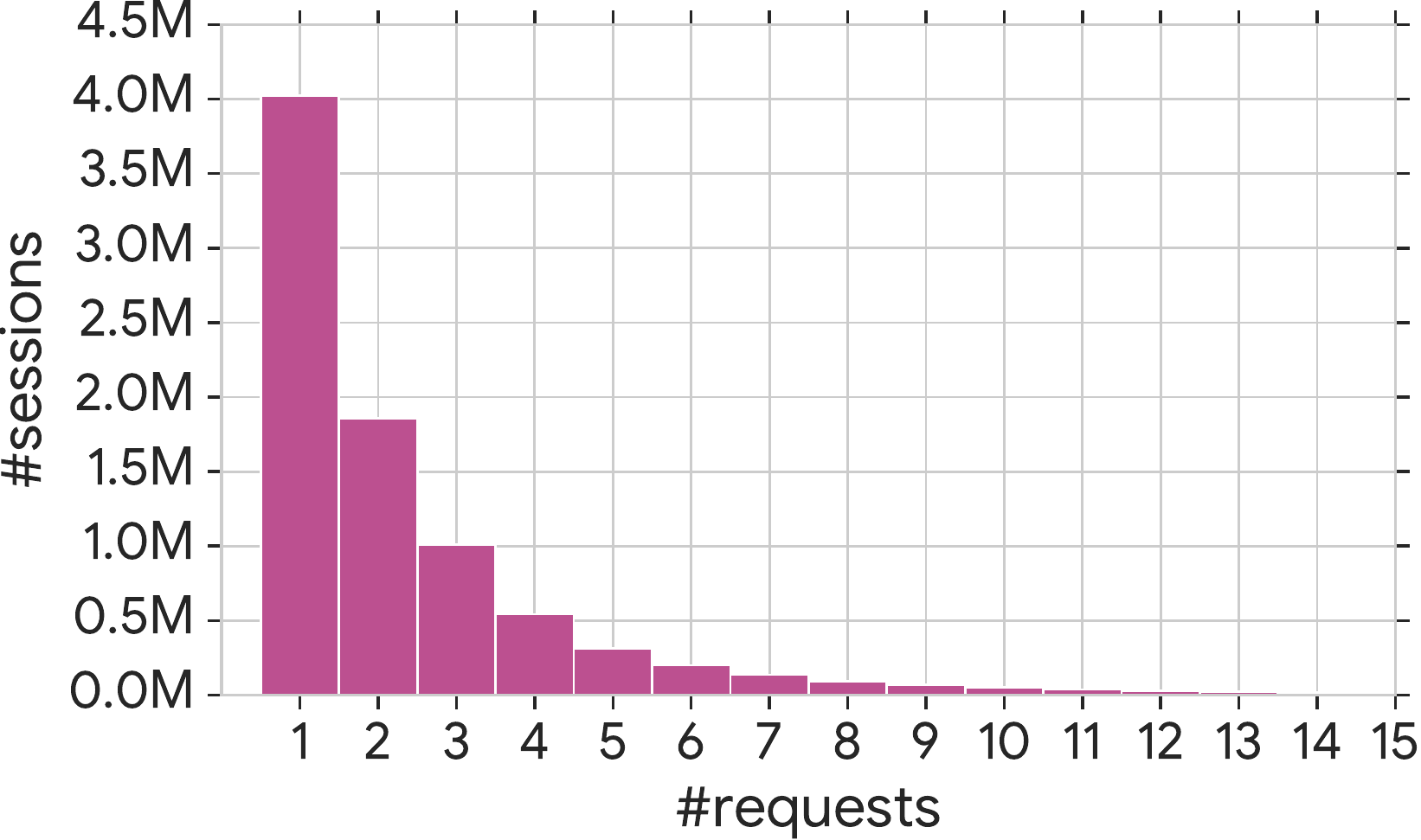}
  \caption{Histogram: number of requests per session}
  \label{fig:requests_per_session}
\end{figure}

\begin{figure}[h]
 \centering
  \includegraphics[width=0.65\linewidth, trim = 0.0cm 0.7cm 0.0cm 0.0cm]{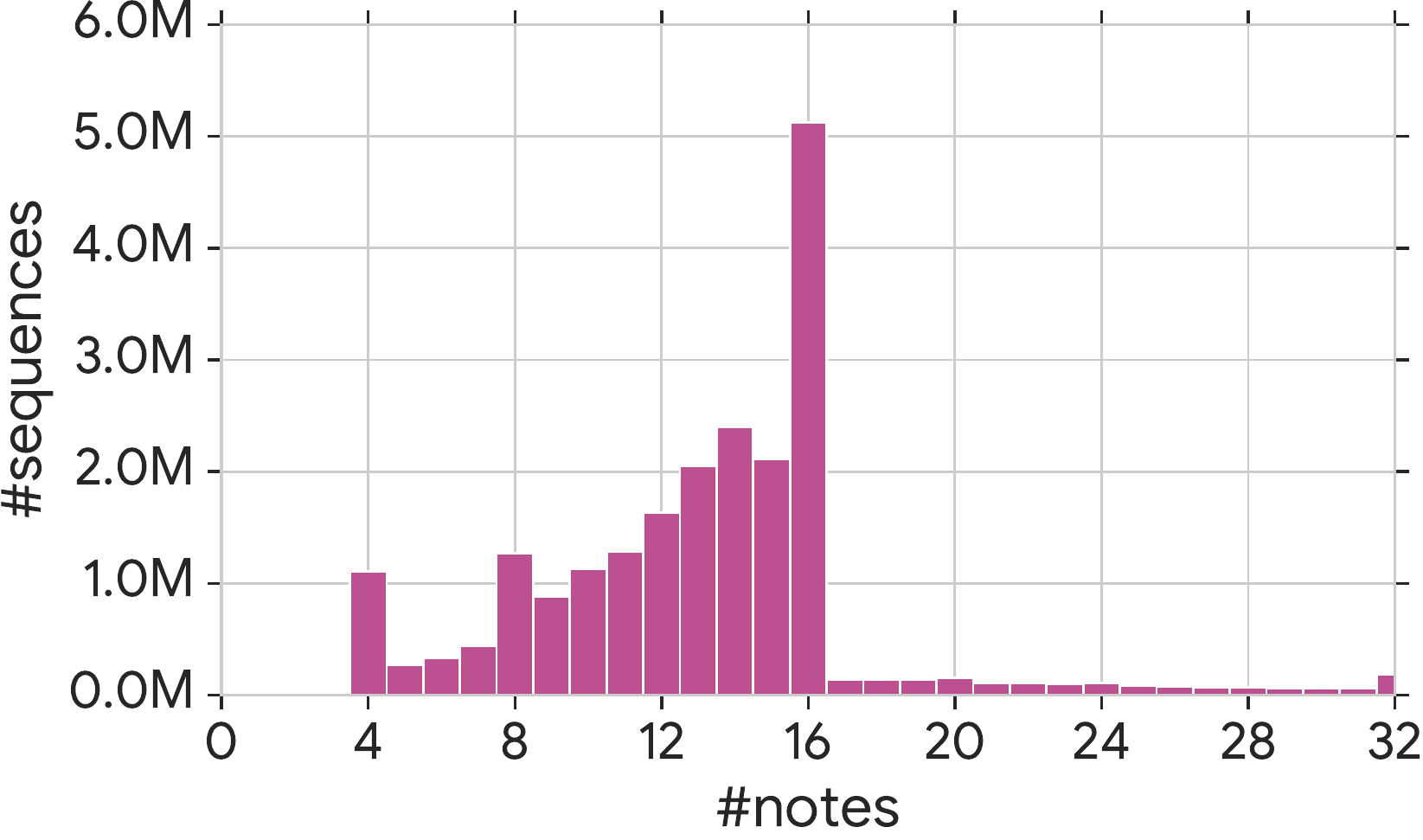}
  \caption{Histogram: length of sequences}
  \label{fig:notes_per_sequence}
\end{figure}

The sequences come from 109 different countries, with the United States, Brazil, Mexico, Italy, and Spain ranking in the top 5.  Countries that had a small number of sequences were all grouped together in a separate category, to minimize the possibility of identifying users.
While many sessions (${\sim}$20\%) contained only one request for harmonization (shown in Figure \ref{fig:requests_per_session}), most sessions had 2 or more harmonizations, either of the same melody, or of a different one. 
As shown in Figure \ref{fig:notes_per_sequence}, more than 5 million of the 
%About 24\% of 
input sequences used the maximum number of notes in the default version of the doodle, which is 16. It is interesting to note that despite being an Easter egg, 7.6\% of user sessions discovered the advanced mode that allowed them to enter longer sequences.
%, as seen in Figure \ref{fig:notes_per_sequence}. 

\begin{figure}[h]
  \includegraphics[width=0.65\linewidth]{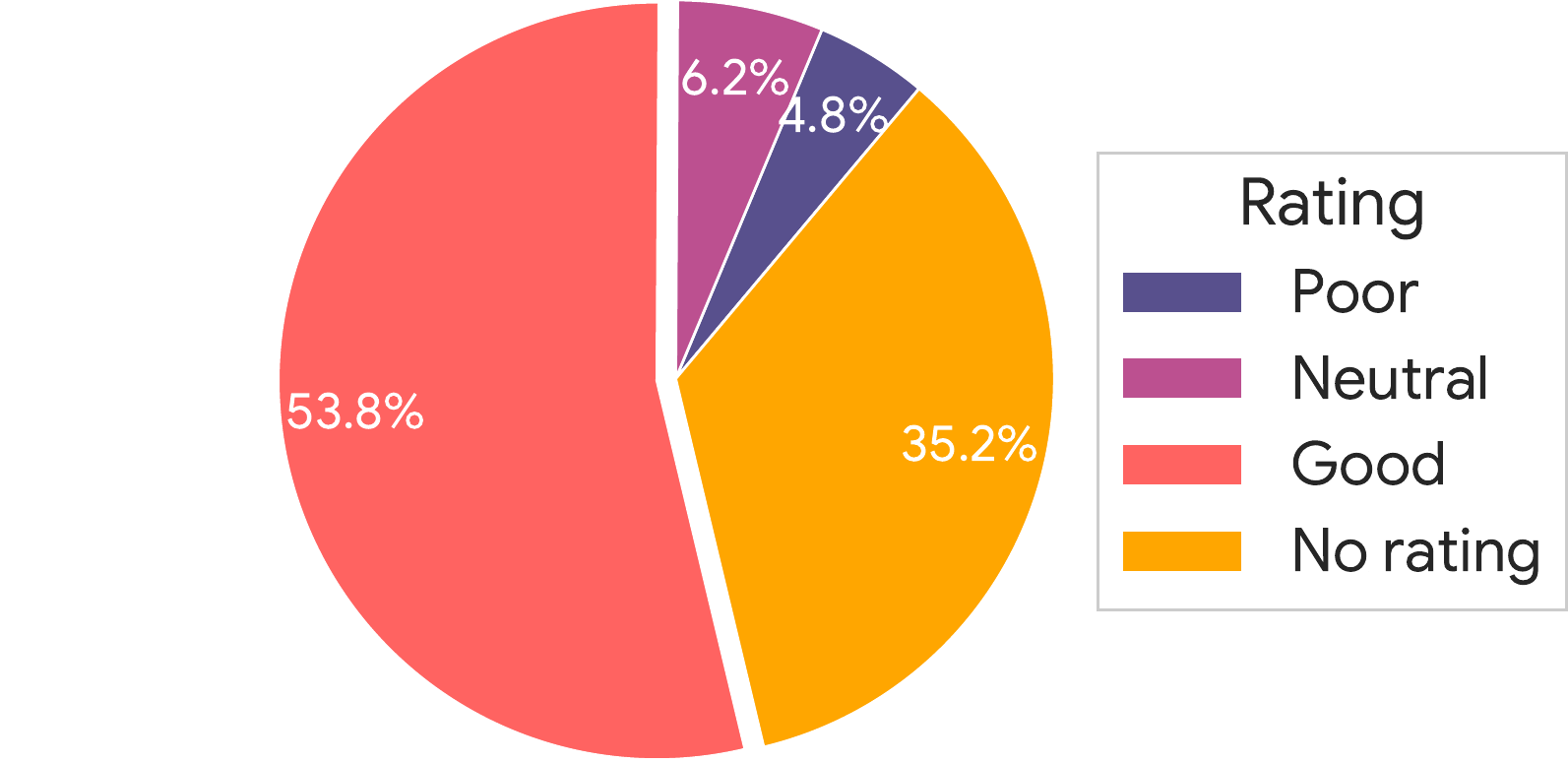}
  \caption{Breakdown of user ratings on harmonized compositions.}
  \label{fig:overall_feedback}
\end{figure}

The doodle has 3 presets: \textit{Twinkle Twinkle Little Star}, \textit{Mary had a Little Lamb}, and the beginning to \textit{Bach's Toccata and Fugue in D Minor, BWV 565}, which are the 3 most repeated sequences.
%, however Figure \ref{fig:top_20_pitches} 
However, there are also shows some surprising runner ups,  
%Along with the presets and sequences where a single note was repeated a number of times, in the top 20 most repeated input melodies we also found 
such as Beethoven's \textit{Ode to Joy}, and \textit{Megalovania}, a popular song from the game \textit{Undertale}, as well as some regional hits~\footnote{Visit {\url{https://g.co/magenta/bach-doodle-dataset}} to interact with visualizations of the top repeated melodies overall and in each region, as well as regional unique hits.}. 
% \begin{figure}[h]
% \centering
%   \includegraphics[width=0.9\linewidth, trim = 0.0cm 0.2cm 0.0cm 0.0cm]{figs/top_20_pitches.pdf}
%   \caption{Most repeated note sequences}
%   \label{fig:top_20_pitches}
% \end{figure}
Overall, users enjoyed their harmonizations, with 53.8\% of all compositions rated as ``Good''.  Figure \ref{fig:overall_feedback} gives the breakdown of user ratings.  
%The entire breakdown of the user feedback can be be seen in Figure \ref{fig:overall_feedback}.

\subsection{Parallel Fifths and Parallel Octaves}

The Coconet model that powered the Bach Doodle was trained to produce harmonizations in the style of Bach chorales, and one well known characteristic of Bach's counterpoint writing is how carefully he followed the rule of avoiding parallel fifths (P5s) and parallel octaves (P8s). However, one complaint from advanced users of the app was the presence of P5s and P8s in the output of the model. Here, we present some analysis of how frequently and under what circumstances such outputs occurred. To identify the P5 and P8 occurrences, we used \emph{music21}~\cite{cuthbert2010music21}.

First, we looked at how frequently P5s and P8s appeared in our training data. We were surprised to find that in the 382 Bach chorale preludes we used in our train and validation sets, there were 132 instances of P5s (0.023/measure) and 51 instances of P8s (0.009/measure). Given this prevalence, the model may learn to output this kind of parallel motion. However, many of these instances can be ``excused'' because they occur under special circumstances such as at phrase boundaries or when using non-chord tones \cite{dahn2016consecutive,fitsioris2008parallel}. Unfortunately, our training data does not include key signatures, time signatures, or fermatas, so the model likely learned to treat P5s as more permissible than was actually the case in Bach's music.

We then examined the output of the model to see if P5s/P8s occurred more frequently when user input was outside the training distribution and if the absence of P5s/P8s was correlated with positive user feedback. We first split the output based on whether users gave positive feedback or not (non-positive feedback includes neutral, negative, and the absence of feedback). Next, we split based on whether user input was within the same pitch range as the soprano lines in the training data (MIDI pitches from 60 through 81) and whether the maximum delta between consecutive pitches exceeded that of the training data (1 octave).

\begin{figure}
  \includegraphics[width=0.95\linewidth,trim = 0.0cm 0.2cm 0.0cm 0.0cm]{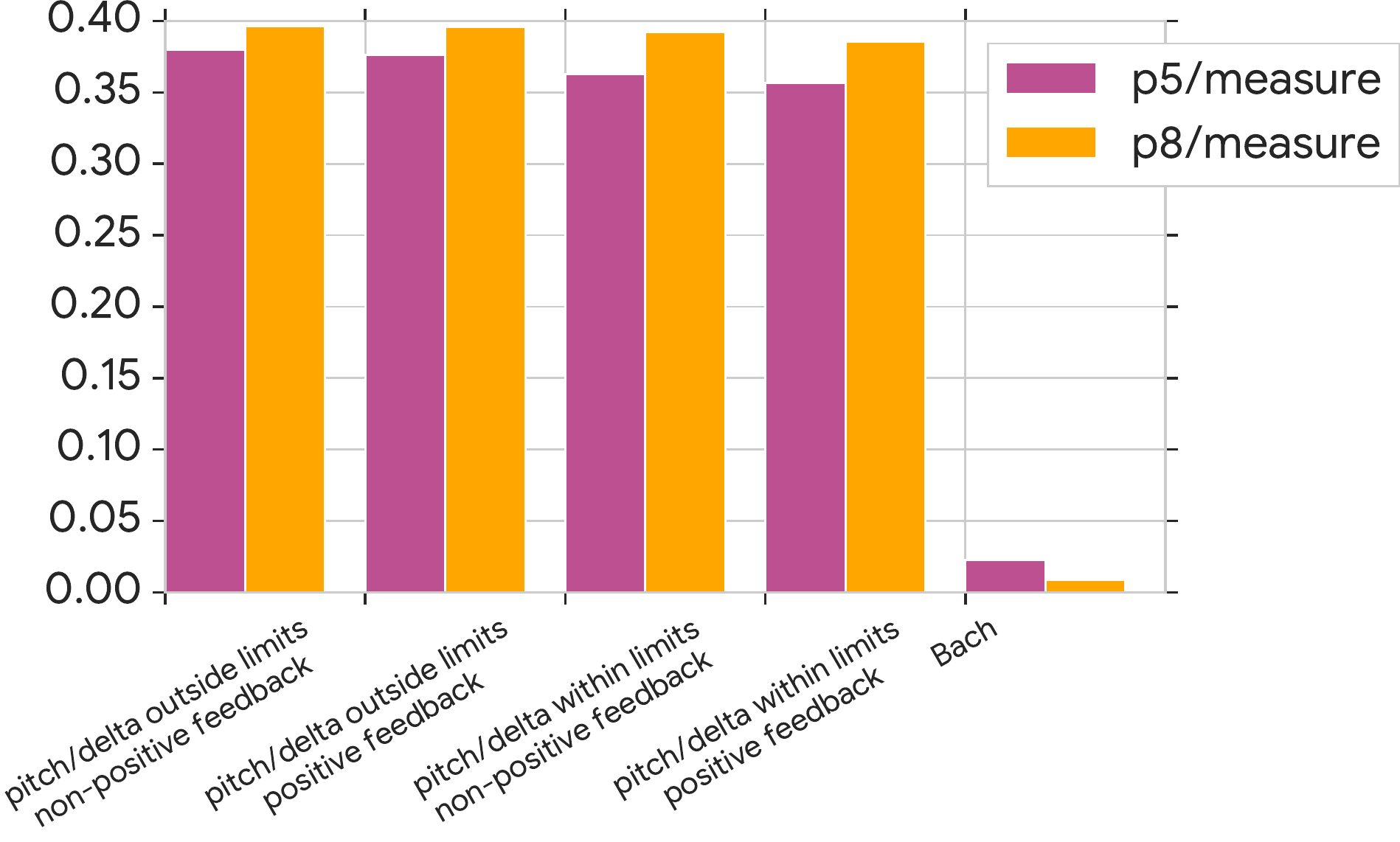}
  \caption{Parallel fifths and octaves per measure}
  \label{fig:parallels}
\end{figure}

In total, we found 15,816,599 P5s (0.365/measure) and 16,949,818 P8s (0.391/measure) in the model output. Results split into the four categories are shown in Figure~\ref{fig:parallels}. As hypothesized, P5s/P8s were more common when user input was out of distribution, and their absence correlated with positive user feedback. A Kruskal-Wallis H test for both the number of P5s and P8s showed that there is at least one statistically significant difference between the four categories with $p < 1\mathrm{e}{-4}$. Further, Mann-Whitney rank tests between the categories showed significant differences, each with $p < 1\mathrm{e}{-4}$.
The correlation between fewer P5s/P8s and positive user feedback is particularly interesting. This could either indicate that users prefer music with fewer P5s/P8s or it could simply mean that when the model produces poor output, P5s/P8s tend to be a feature of that output. In any case, the presence of P5s/P8s seems to be a useful proxy metric for model output quality. In future work, it could be a useful signal during training (similar to \cite{jaques2017sequence}), evaluation, or perhaps even during inference where it could trigger additional Gibbs sampling steps.

\section{Conclusion}
%We provided a case study of how we launched a popular web app that uses machine learning to make music composition more accessible.  
%
The Bach Doodle enabled large-scale participation in baroque-style counterpoint composition through an intuitive sheet music interface assisted by machine learning. We hope this encourages more creative apps that allow novices and artists to interact with music composition and machine learning in approachable ways. 
%Our composition app enabled an unprecedented amount of enthusiasm around the globe in classical music, in music composition and in composing with machine learning.  We hope this encourages more creative apps that allow novices and artists to interact with machine learning in approachable ways.   
With this paper, we are releasing a dataset of 21.6 million instances of human-computer collaborative miniature compositions, along with meta-data such as user rating and country of origin.  We hope the community will find it useful for ethnomusicological studies, music education, or improving machine learning models.

\section{Acknowledgements}
Many thanks to Ann Yuan, Daniel Smilkov and Nikhil Thorat from Tensorflow.js for their expert assistance.
A big shoutout to Pedro Vergani, Rebecca Thomas, Jordan Thompson and others on the Doodle team for their contribution to the core components of the Doodle.  Thank you Lauren Hannah-Murphy and Chris Han for keeping us on track.  Thank you to Magenta colleagues for their support. Thank you Tim Cooijmans for co-authoring the blog post.

% For bibtex users:
\bibliography{bach}

\end{document}